\documentclass[twocolumn,showpacs,prb,epsfig]{revtex4}
\usepackage{amsmath,amssymb}
\usepackage{graphicx,bm}
\usepackage{epsfig}
\newcommand{\Tr}{{\rm Tr}}

\begin{document}
\def\be{\begin{equation}}
\def\ee{\end{equation}}
\def\Tr{{\rm Tr}}

\title{Effect of a Zeeman field on the superconductor-ferromagnet transition in metallic grains}
\author{S. Schmidt}
\author{Y. Alhassid}
\affiliation{Center for Theoretical Physics, Sloane Physics Laboratory, Yale University, New Haven, Connecticut 06520, USA}
\author{K. Van Houcke}
\affiliation{Universiteit Gent, Vakgroep Subatomaire en Stralingsfysica - Proeftuinstraat 86, B-9000 Gent, Belgium}
\begin{abstract}
We investigate the competition between pairing correlations and ferromagnetism in small metallic grains in the presence of a Zeeman field. Our analysis is based on the universal Hamiltonian, valid in the limit of large Thouless conductance. We show that the coexistence regime of superconducting and ferromagnetic correlations can be made
experimentally accessible by tuning an external Zeeman field.  We compare the exact solution of the model with a mean-field theory and find that the latter cannot describe pairing correlations in the intermediate regime. We also study the occurrence of spin jumps across the phase boundary separating the superconducting and coexistence regimes.
\end{abstract}

\maketitle
\section{Introduction}
The hallmark of the BCS model of
superconductivity in metals is the presence of an excitation gap
$\Delta$. This gap is caused by the formation of Cooper pairs describing correlated electron pairs in time-reversed
states. Thus, pairing correlations in superconductors tend to
minimize the total spin of the electron system.
Ferromagnetic correlations, on the other hand,  prefer to maximize the total spin
and form a macroscopic magnetic moment. Early work~\cite{abri,clog,chan,fuld,lark} predicted a state in which both pairing and ferromagnetic order are present, if  ferromagnetism is caused by localized paramagnetic impurities.
The experimental observation that both states of matter can
coexist in heavy fermion systems~\cite{sax,pfei,aoki} and high-Tc
superconductors~\cite{tall,bern} came as a surprise and led to the search for
new theoretical models to describe this coexistence. 
A BCS-like model of $s$-wave pairing combined with a simple Stoner-like model of ferromagnetism was used to derive such an intermediate state within a mean-field approximation.~\cite{karchev} However, it was argued that such a state is unstable in the bulk.~\cite{zhou, shen01, shen02} Furthermore, it was shown that a proper Hartree-Fock mean-field theory of the model does not support coexistence of $s$-wave superconductivity and ferromagnetism.~\cite{ jogle, blagoev} 

A similar model of BCS-like pairing and exchange interaction was shown to be valid in small metallic grains in the mesoscopic regime for a Thouless energy $E_T$ that is large compared with the single-particle mean-level spacing $\delta$.~\cite{kur,alei,murthy01} In such a finite-size system, a partially paired state with finite spin polarization exists within a narrow parameter regime.~\cite{ying} Since this coexistence regime is relatively small, it would be difficult to observe it experimentally. It has been suggested that the probability of spin polarization in the presence of pairing correlation may be enhanced by mesoscopic fluctuations~\cite{falci} or by an asymmetric spin-dependent bandwidth of the single-particle spectrum.~\cite{ying} 

Here we study the competition between ferromagnetic and pairing correlations in metallic grains in the crossover regime from a few-electron system ($\Delta\ll\delta$) to the bulk ($\Delta\gg\delta$). We use Richardson's solution of the BCS-like interaction~\cite{rich} and the known solution of the exchange model~\cite{rupp} to determine the ground state of the grain.  For sufficiently small grains, there is a regime in pairing gap $\Delta/\delta$ and exchange coupling $J_s/\delta$, in which the ground state is partly paired and partly polarized.  We show that, in the presence of a Zeeman field, the exchange coupling at which the crossover from a pure superconducting state to the coexistence regime takes place decreases to values that can be realized in several metals. The onset of magnetization with increasing exchange coupling at a given pairing gap corresponds to a spin jump $\Delta S \geq 1$, followed by successive spin increments of $\Delta S=1$. The magnitude of the initial spin jump depends on the value of $\Delta/\delta$.  Similar spin jumps were found in the crossover from a superconducting state to a paramagnetic state,~\cite{delft01,delft02} where they are reminiscent of a first-order transition in the bulk. We apply a mean-field theory similar to the one used in Ref.~\onlinecite{delft02} and compare with the exact results. In contrast to the exact solution, we find that the mean-field approximation cannot describe pairing correlations in the intermediate regime of partial spin polarization.
\section{Model}
An isolated metallic grain in which the single-particle dynamics are chaotic and whose dimensionless Thouless conductance $g_T=E_T/\delta$ is large ($g_T \gg 1$), can be described by an effective universal Hamiltonian~\cite{kur,alei,murthy01}
\begin{eqnarray}
\label{origH}
\hat H=\sum_{k\sigma}\hspace{-0.0cm}\epsilon_{k} c_{k\sigma}^\dagger c_{k\sigma}
 -G \hat P^\dagger \hat P  -  J_s \hat{\bf S}^2  +  g \mu_B H \hat S_z\;.
\end{eqnarray}
Here $c_{k\sigma}^\dagger$ is the creation operator for an
electron in the single-particle level $\epsilon_k$ with either spin up
($\sigma=+$) or spin down ($\sigma=-$). The one-body term in (\ref{origH}) describes
the kinetic energy plus confining single-particle potential. The second term on the r.h.s. of Eq.~(\ref{origH}) is a pairing interaction with strength $G$ and where $P^\dagger=\sum_i c_{i+}^\dagger c_{i-}^\dagger$ is the pair creation operator. The third term in (\ref{origH}) is an exchange interaction expressed in terms of the
total spin operator $\hat{\bf S}=\sum_{k\sigma\sigma'}c_{k\sigma}^\dagger {\bf \tau}_{\sigma\sigma'}c_{k\sigma'}$ ($\tau_i$ are Pauli matrices). The parameter $J_s$ is the exchange coupling constant (estimated values of $J_s$ for  a variety of materials were tabulated in Ref.~\onlinecite{gorok}). The inclusion of such an exchange interaction in quantum dots~\cite{al00} explained quantitatively the measured peak height and peak spacing statistics.~\cite{mal,rupp} The last term on the r.h.s. of Eq.~(\ref{origH}) describes the coupling of an external Zeeman field $H$ (applied in the $z$ direction) to the spin of the dot. Here $g$ is the $g$-factor of the electrons in the grain (taken to be positive) and $\mu_B$ is the Bohr magneton. Orbital diamagnetism can be neglected for small grains.~\cite{delft02} The charging energy $e^2 \hat N^2/2C$ ($C$ is the capacitance of the grain) is a constant for a grain with a fixed number of electrons $N$ and was omitted in the Hamiltonian (\ref{origH}).

In this work, we do not consider mesoscopic effects that originate in the random matrix description of the single-particle Hamiltonian.  To construct a typical phase diagram of a single grain, we consider a generic equidistant spectrum $\epsilon_{k}=k\delta$ with $-N_o\leq k \leq N_o$ at half filling. Thus we have $N=2N_o$ for an even number of electrons ($p=0$) and $N=2N_o+1$ for an odd number ($p=1$).
\section{Exact solution}
In the absence of a pairing interaction ($G=0$), the Hamiltonian (\ref{origH}) can be solved in closed form.~\cite{rupp} The orbital occupations $\hat n_k=\hat n_{k+}+ \hat n_{k-}$ commute with $\hat {\bf S}^2$ and are good quantum numbers. The empty ($n_k=0$) and doubly occupied ($n_k=2$) orbitals do not contribute to the total spin, so the total spin of the grain is obtained by coupling the singly occupied levels with spin $1/2$ to total spin $S$ and spin projection $M$. For a specific set $\cal{B}$ of $b$ singly occupied levels, the total spin ranges from $S=p/2$ to $S=b/2$ with each spin value having a degeneracy of $d_b(S) = \binom{b}{S+ b/2} - \binom{b}{S+1+ b/2}$. A complete set of eigenstates is then given by $\vert \{n_k\}, \gamma, S, M\rangle $ where $\gamma$ are quantum number distinguishing between eigenstates with the same spin.~\cite{rupp,tureci}

The pairing interaction can only scatter time-reversed pairs from doubly occupied to empty orbitals but does not affect the singly occupied levels (referred to as ``blocked'' levels). It is therefore sufficient to diagonalize the reduced BCS Hamiltonian $\sum_{k \sigma}\hspace{-0.0cm}\epsilon_k  c_{k\sigma}^\dagger c_{k\sigma} -G P^\dagger P$ using the single-particle subspace ${\cal U}$ of empty and doubly occupied levels. This problem was solved by Richardson.~\cite{rich}  The eigenenergies are given by
\begin{eqnarray}
E_m=\sum_{\mu=1}^m E_\mu\;,
\end{eqnarray}
where $E_\mu$ are parameters that characterize the eigenstate and $m=(N-b)/2$ is the number of pairs. Richardson's parameters $E_\mu$ are found by solving the set of $m$ coupled non-linear equations \begin{eqnarray}\label{richardson}
\frac{1}{G}+2\sum_{\nu=1\atop \nu\neq\mu}^{m}\frac{1}{E_\nu-E_\mu}=
\sum_{i\in\cal{U}}\frac{1}{2\epsilon_i-E_\mu} \;\;\;\;(\mu=1,\ldots,m)\;.
\end{eqnarray}
To each set of $m$ doubly occupied levels at $G=0$, there is a unique solution for Richardson's parameters at $G \neq 0$.  We note that in the general case Richardson's equations depend on the seniority quantum numbers of the levels (the seniority is the number of electrons not coupled to spin zero). In our case, the levels are doubly degenerate and the seniority of a doubly occupied level is zero.

The eigenstates constructed from the subset ${\cal U}$ of empty and doubly occupied levels have spin zero, so the total spin of the grain is determined by the spin-1/2 coupling of the singly occupied levels in ${\cal B}$.
The eigenstates of the full Hamiltonian (\ref{origH}) are then given by $\vert {\cal B},\{ E_\mu\}, \gamma, S,M \rangle$ with energies of
\begin{eqnarray}\label{Eexact}
E =E_m + \sum_{k\in \cal{B}}\epsilon_k - J_s S (S+1)+ g\mu_B H M \,.
\end{eqnarray}

In this work, we focus on the ground state of the grain as a function of the interaction couplings $G$ and $J_s$. To that end, we find the lowest energy $E(S)$ in (\ref{Eexact}) for a given spin $S$ and then minimize with respect to $S$. The energy $E(S)$ is found by choosing a set ${\cal B}$ of $b=2S$ singly occupied levels that are placed
closest to the Fermi energy. We then populate these $b$ levels with spin-down electrons, resulting in a good-spin state with spin $S$ and spin projection $M=-S$. For a given set ${\cal B}$, we solved Richardson's equations using the method of Ref.~\onlinecite{rombouts}.

The physical parameter desc
ribing the pairing Hamiltonian is $\Delta/\delta$, where $\Delta$ is the bulk pairing gap and $\delta$ the single-particle mean-level spacing. The low-energy spectrum of the grain (for $J_s=H=0$) is determined by the value of this parameter. We can truncate the total number of levels from $N_o$ to $N_r < N_o$,  and renormalize $G$ such that the low-energy spectrum of the grain remains approximately the same. For a picketfence spectrum, the renormalized coupling constant is given by
\begin{eqnarray}\label{renorm}
\frac{G_r}{\delta}=\frac{1}{{\rm arcsinh}\left(\frac{N_r+1/2}{\Delta/\delta}\right)}\;.
\end{eqnarray}
Strictly speaking, this holds in the absence of an exchange interaction. However, since the exchange interaction affects only the blocked levels, we expect the renormalization (\ref{renorm}) to hold as long as the number of blocked levels is small compared with the total number of levels in the band. The quality of this approximation depends on the choice for $N_r$ and was discussed in detail in Ref.~\onlinecite{al02}.
\section{Mean-field approximation}
We compare the findings based on the exact solution with a mean-field theory. The mean-field approach is based on a trial wave function of the form~\cite{delft02}
\begin{eqnarray}
\vert\psi_S\rangle=\prod_{k\in {\cal B}}c^\dagger_{k-}\prod_{j\in {\cal U}}\left(u_j^{(S)}+v_j^{(S)}c_{j+}^\dagger c_{j-}^\dagger\right)\vert0\rangle
\end{eqnarray}
with the normalization condition $(u_j^{(S)})^2+(v_j^{(S)})^2=1$. The wave function $\psi_S$ has $b=2S$ singly occupied levels with spin-down electrons (set ${\cal B}$) chosen to be closest to the Fermi energy, and is of the BCS form within the remaining set of levels ${\cal U}$. The lowest state with spin $S$ is found by minimizing
the expectation value $\langle\psi_S\vert (\hat{H}-\mu \hat{N})\vert\psi_S\rangle$ with respect to the variational parameters $v_j^{(S)}$. Here $\mu$ is a chemical potential ensuring that the average number of particles is $N$.

The mean-field energy at fixed spin $S$ is given by
\begin{eqnarray}
\label{Evar}
E_{\rm mf}\left(S\right)&=& 2\sum_{k\in {\cal U}}\hspace{-0.1cm}\epsilon_k \left(v_k^{(S)}\right)^2-\frac{\Delta_S^2}{G}\nonumber\\
&+&\sum_{k\in {\cal B}}\hspace{-0.1cm}\epsilon_k-J_s S\left(S+1\right)-g \mu_B H S\,,
\end{eqnarray}
where
\begin{eqnarray}
\left(v_k^{(S)}\right)^2=\frac{1}{2}\left(1-\frac{\epsilon_k-\mu}
{\sqrt{\left(\epsilon_k-\mu\right)^2+\Delta_S^2}}\right)\;,
\end{eqnarray}
and $\Delta_S$ is a spin-dependent pairing gap. The  gap parameter and chemical potential are determined by solving the
gap equation together with the particle number equation
\begin{subequations}
\begin{eqnarray}
\label{g}
\frac{2}{G} & = & \sum_{k\in {\cal U}}\frac{1}{\sqrt{\left(\epsilon_k-\mu\right)^2+\Delta_S^2}} \;,\\
N & = & 2\sum_{k\in {\cal U}}\hspace{-0.1cm}\left(v_k^{(S)}\right)^2+ b \;.
\end{eqnarray}
\end{subequations}
For an equidistant spectrum, the chemical potential can be determined by symmetry considerations and is given by $\mu=-(1-p)\delta/2$ for $N_o\delta\gg\Delta_S$.

Here, we used the same approximations as in Ref.~\onlinecite{delft02} and neglected a term in the energy $E_{\rm mf}(S)$ which is proportional to $(v_j^{(S)})^4$. The result (\ref{Evar}) is in agreement with the leading term of an expansion in the inverse number of electrons $1/N$.~\cite{yuzba}. Comparing (\ref{Evar}) and (\ref{Eexact}) with $M=-S$, we see that the exchange and Zeeman terms are treated exactly in this mean-field approximation. The ground-state spin in the mean-field approximation is found by minimizing $E_{\rm mf}(S)$ in (\ref{Evar}) with respect to $S$.

\section{Ground-state phase diagram}
The ground-state spin of the grain is determined by the competition between various terms in the universal Hamiltonian. The one-body part (kinetic plus confining one-body potential) and pairing interaction favor minimal spin $S=p/2$ while exchange interaction and Zeeman field favor a maximally polarized state. We have studied the ground-state spin as a function of the three parameters $\Delta/\delta , J_s/\delta$ and $g\mu_B H/\delta$. Using the exact solution, we find three different phases: a superconducting phase where the number of pairs is maximal and $S=p/2$, a ferromagnetic phase where the system is fully polarized $S=N/2$ (all electrons are with spin down), and an intermediate regime. The intermediate regime describes a partially polarized state $S< N/2$, in which $b=2S$ electrons reside in singly occupied levels closest to the Fermi energy and the remaining electrons are paired to give spin zero.

\begin{figure}[t]
\epsfxsize=0.95\columnwidth{\epsfbox{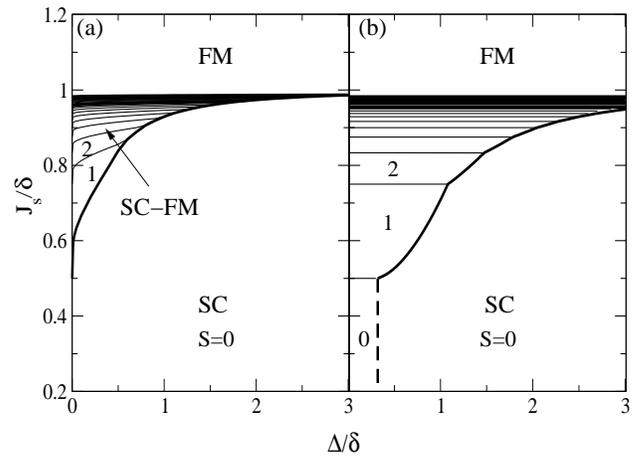}} \caption{Ground-state phase diagram in the $J_s/\delta$--$\Delta/\delta$ plane for an even number of electrons ($N=50$). Left panel: exact results. Right panel: mean-field approximation (see text). The phase diagrams show a superconducting (SC) phase and a ferromagnetic (FM) phase. The exact phase diagram also exhibits an intermediate regime (SC-FM) in which the ground state is partially polarized but still has pairing correlations.  The intermediate regime in the mean-field phase diagram describes a state that is partially polarized state but does not include pairing correlations. In particular, the dashed line separates an $S=0$ SC phase from an $S=0$ phase with no pairing correlations. The numbers shown in the intermediate regime are the spin values in the corresponding sectors.}\vspace{-0.5cm}
\end{figure}
The phase diagram in the $\Delta/\delta$--$J_s/\delta$ plane of a grain with even number of electrons and in the absence of Zeeman field ($H=0$) is presented in Fig.~1(a). For weak pairing, the superconducting and ferromagnetic phases are separated by an intermediate regime. The boundaries of this intermediate regime are described by two critical values $J_s^{(1)}$ and $J_s^{(2)}$ of the exchange interaction that are function of $\Delta/\delta$. The critical value $J_s^{(1)}/\delta$ is a monotonically increasing function of $\Delta/\delta$, i.e., a stronger exchange interaction is required to polarize a grain with stronger pairing correlations.  However, $J_s^{(2)}/\delta$ is almost insensitive to $\Delta/\delta$.  The intermediate regime shrinks at larger $\Delta/\delta$ and eventually disappears above $\Delta/\delta \sim 3$. For stronger pairing correlations, the superconducting phase makes a direct transition to the ferromagnetic phase. In this regime (not shown in Fig.~1), the phase boundary exhibits a strong dependence on the bandwidth $N_o$.

For comparison, we show the mean-field results in Fig.~1(b). We observe that the mean-field results are qualitatively different from the exact solution. The region to the right of the thick solid line and dashed line describes an superconducting phase with $\Delta _0 \neq 0$. However, there is no superconducting solution (i.e., $\Delta_0=0$) for $\Delta/\delta \leq 0.28$. Furthermore, in each of the partially polarized regions with spin $0<S<N/2$, the corresponding pairing gap vanishes  $\Delta_S=0$ and there are no pairing correlations present. While solutions with $\Delta_S \neq 0$ exist, they occur for values of $\Delta/\delta$ for which a higher spin state with no pairing correlations has lower energy (because of the exchange interaction). For example, a solution with $\Delta_1 \neq 0$ exists only for $\Delta/\delta > 2.1.$~\cite{delft02} However, at this strength of pairing correlation we observe a direct transition from $S=0$ to $S=4$ with $\Delta_S=0$ as the energy of the lowest $S=4$ state with no pairing correlations is lower than the paired $S=1$ state.  As a result, the boundaries which separate different spin phases are flat, e.g., independent of the pairing strength.  In fact, in the mean-field approximation the ground-state wave function is a Slater determinant through the whole intermediate regime of partial spin polarization. Thus no coexistence of pairing and spin polarization is observed within the mean-field approach.

In contrast, the exact solution shows that pairing correlations are present as long as the system is not fully polarized. This can be seen in the shift of the spin transition lines to higher values of the exchange interaction strength as the pairing gap $\Delta/\delta$ is increased. Thus, the exact solution predicts a regime in which pairing correlations and spin polarization coexist. In the following, we only discuss results obtained from the exact solution.

More detailed phase diagrams for $H=0$ are shown in the top row of Fig.~2 for both grains with even [panel (a)] and odd [panel (b)] number of electrons. For weak pairing we observe an odd-even effect (in number of electrons). In particular, the critical value $J_s^{(1)}$ is larger for the odd grain, even though the presence of a blocked level in the odd superconducting phase weakens pairing correlations in the odd grain. This is because increasing the spin from $1/2$ to $3/2$ in the odd grain costs more one-body energy than increasing the spin from $0$ to $1$ in the even grain.

The phase boundaries for a finite Zeeman field $g\mu_B H/\delta=2.6$ are shown in the bottom row of Fig.~2. The effect of a Zeeman field is twofold. First, it helps polarizing the grain, making the value of $J_s^{(1)}$ for a given pairing gap smaller. Second, at given exchange strength $J_s/\delta$, it increases the critical value of $\Delta/\delta$ at which partial spin polarization is destroyed. Both effects together increase the size of the intermediate regime in the $\Delta/\delta$--$J_s/\delta$ plane.
\begin{figure}[t]\label{fig2}
\epsfxsize=1.0\columnwidth{\epsfbox{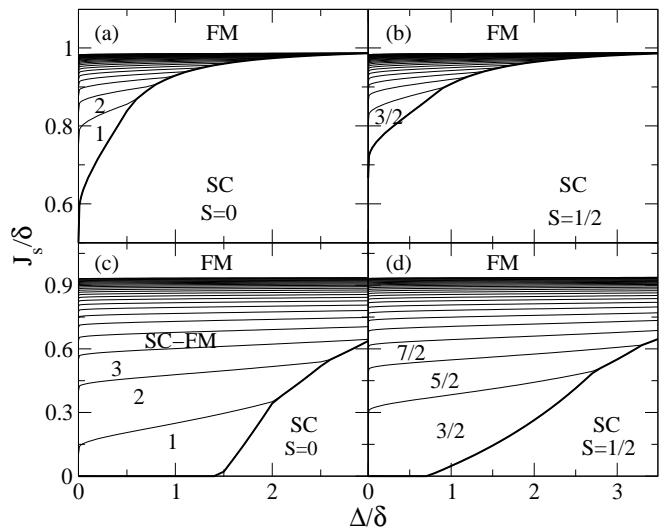}} \caption{Phase diagrams in the $J_s/\delta$--$\Delta/\delta$ plane at a fixed Zeeman field $g\mu_B H=0$ (top panels) and $g\mu_B H/\delta=2.6$ (bottom panels) for an even grain (left panels) and for an odd grain (right panels). The numbers denote the spin in each sector.}
\end{figure}
\section{Spin jumps}
As we increase the exchange coupling constant $J_s/\delta$ at fixed  $\Delta/\delta$ and Zeeman field, the spin increases by discrete steps from its minimal value $S=p/2$ to its maximal value of $S=N/2$. In the absence of pairing ($\Delta=0$), the transition from spin $S$ to spin $S+1$ occurs for an exchange coupling of
\begin{eqnarray}
J_s/\delta=\frac{\left(2S+1\right)-g\mu_B H/\delta}{2S + 2}\qquad\mbox{at}\quad \Delta=0 \;.
\end{eqnarray}
\begin{figure}[t]\label{fig3}
\epsfxsize=1.0\columnwidth{\epsfbox{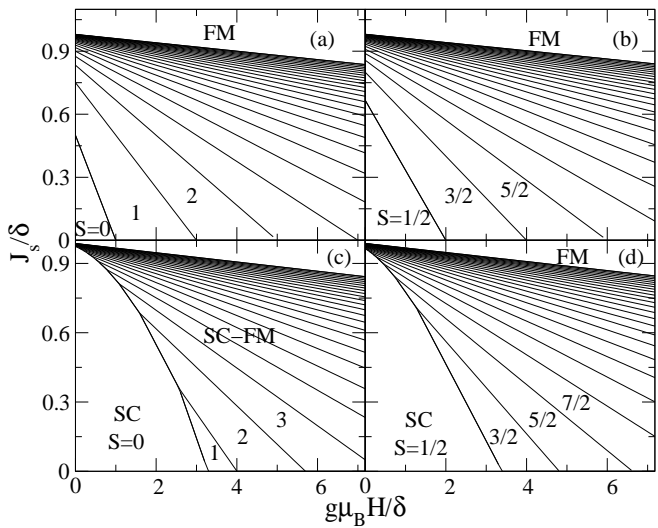}} \caption{Phase diagrams in the $J_s/\delta$--$g \mu_B H/\delta$ plane at fixed  $\Delta/\delta=0$ (top row) and $\Delta/\delta =2$ (bottom row) for an even grain (left column) and for an odd grain (right column). Numbers denote the spin in each sector.}
\end{figure}
The corresponding $\Delta=0$ phase diagrams in the $g\mu_BH/\delta$--$\Delta/\delta$ plane are shown in Fig.~3(a) and 3(b) for even and odd grains, respectively.  In particular, the phase boundaries are given by $J_s^{(1)}=\delta(p+1)/(p+2)-g\mu_B H/(p+2)$ and $J_s^{(2)}=\delta(N -1)/N -g\mu_B H/N$.  The ground-state spin increases as a function of $J_s$ in steps of $\Delta S=1$.
An interesting qualitative change in the presence of pairing correlations is the possibility of a spin jump $\Delta S >1$.  For $\Delta/\delta < 0.6$, the ground-state spin still increases in steps of $\Delta S=1$ versus $J_s$.  However, for  $0.6 < \Delta/\delta < 0.8$, the ground-state spin jumps from $0$ to $2$ within the range $0.87< J_s/\delta < 0.9$.  The size of the first-step spin jump gets larger with increasing $\Delta/\delta$. All subsequent steps are of size one [see Fig.~2(a)].

A similar effect was observed when superconductivity breaks down due to the presence of a large external Zeeman field.~\cite{delft02} The experimental findings were qualitatively explained using the mean-field theory we discussed previously (but without the inclusion of an exchange interaction). It was concluded in Ref.~\onlinecite{delft02} that the first-order phase transition from a superconductor to a paramagnet, observed in thin films, is ``softened'' in metallic grains.
Here we have shown that spin jumps also occur in the presence of exchange correlations. In the absence to an external Zeeman field, these spin jumps are predicted to occur at exchange coupling values $J_s/\delta> 0.87$. Such exchange coupling values are significantly larger than the values for most metals (see Fig.~9 in Ref.~\onlinecite{gorok}). Moreover, the exchange is an intrinsic material property and is difficult to tune experimentally.

The regime of spin jumps can be tuned to lower and more typical values of $J_s$ by applying an external Zeeman field. We have already seen in Figs.~2(c) and 2(d) that a relatively weak Zeeman field increases the size of the intermediate regime. It also means that spin jumps can be observed at smaller values of the exchange strength that are accessible to experiments. This is demonstrated in Fig.~3(c) and 3(d) where phase diagrams in the $g\mu_BH/\delta$--$\Delta/\delta$ plane are shown for a given pairing gap of $\Delta/\delta=2$. For example, a Zeeman field of $g\mu_B H/\delta \approx 2$ is sufficient to lower the critical exchange strength for the $0 \to 2$ spin jump to $J_s/\delta\approx 0.55$ at $\Delta/\delta=2$ [see Fig.~3(c)] as compared to $J_s=0.89\delta$ at $\Delta=0.7\delta$ without Zeeman field [see Fig.~2(a)].

The idea of a Zeeman-field tuning of the values of exchange coupling where spin jumps occur is best illustrated in Fig.~4, where spin staircase functions are shown versus $J_s/\delta$. In the presence of pairing correlations and the absence of Zeeman field, the ground-state spin staircase is shifted to the right and compressed as $\Delta/\delta$ increases [see Fig.~4(a)], reflecting the fact that the intermediate region shrinks [see Fig.~2(a) and 2(b)]. For an even grain with $\Delta/\delta=0.7$, a spin jump of $\Delta S=2$ sets in at $J_s\approx 0.89\delta$, while for $\Delta/\delta=0.9$, a spin jump of  $\Delta S=3$  occurs at $J_s\approx 0.92\delta$. For a finite Zeeman field of $g\mu_B H/\delta = 2$, the spin staircase functions that exhibit similar spin jumps are shifted to smaller values of the exchange strength but larger values of the pairing gap [see Fig.~4(b)]. Spin jumps of $\Delta S=2$ ($\Delta S=3$) occur at $J_s/\delta=0.55$ ($J_s/\delta=0.64$) and $\Delta/\delta =2$ ($\Delta/\delta =2.3$).

In the relevant experimental situation of a fixed exchange interaction strength, the critical value of $\Delta/\delta$ at which spin jumps occur as well as the size of these jumps increase at larger values of $g\mu_B H/\delta$. The ratio $\Delta/\delta$ (for the given metal) can be made larger by studying a larger grain, hence reducing the mean level spacing $\delta$.

As an example, niobium has an exchange interaction strength of $J_s/\delta \approx 0.4$ (see, e.g., Fig. 9 in Ref.~\onlinecite{gorok}). Without an external Zeeman field, the ground-state spin will be minimal ($S=p/2$) at all values of $\Delta/\delta$ [see Fig. 2(a) and 2(b)]. At a Zeeman field $g\mu_B H/\delta = 1$, the ground-state spin of niobium changes from 0 to 1 at $\Delta/\delta =0.66$. However, at $g\mu_B H/ \delta=2.6$, a spin jump of $\Delta S=2$ occurs from 0 to 2 at $\Delta/\delta=2.15$ [see Fig. 2(c)]. For these two values of $\Delta/\delta$, we can roughly estimate the corresponding critical size of the metallic grain given the bulk gap value $ \Delta=3.05 \,{\rm meV}$ and Fermi momentum $ k_{\rm F}=11.8 \,{\rm nm}^{-1} $ of niobium. In a Fermi gas model, the mean-level spacing is related to the volume of the grain by $\delta=2\pi^2\hbar^2/(m k_{\rm F} V)$. Assuming a hemispheric grain with radius $r$, we have the relation $r_{\rm Nb}\approx 2.7\, {\rm nm} \left(\Delta_{\rm Nb}/\delta\right)^{1/3}$. Thus, the Hamiltonian (\ref{origH}) with an equidistant spectrum predicts a $0 \to 1$ spin transition for a niobium grain of radius $r\approx 2.35\, {\rm nm}$ and Zeeman field of $g\mu_B H=4.62\, {\rm meV}$, and a $0 \to 2$ spin jump at $r\approx 3.48 \,{\rm nm}$ and a Zeeman field of $g\mu_B H=3.69 \,{\rm meV}$.
\begin{figure}[t]\label{fig4}
\epsfxsize=0.95\columnwidth{\epsfbox{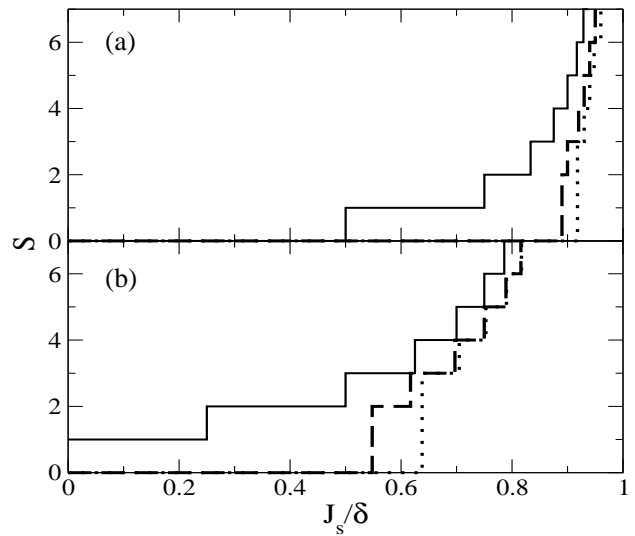}} \caption{Ground-state spin versus exchange coupling $J_s/\delta$ for an even grain at a fixed Zeeman field strength $g\mu_B H=0$ (top panel) and $g\mu_B H/\delta=2$ (bottom panel). Solid lines correspond to a grain with no pairing correlations $\Delta/\delta=0$. The dashed lines describe staircase functions with a spin jump of $\Delta S=2$ for $\Delta/\delta=0.7$ (top) and $\Delta/\delta=2$ (bottom). The dotted lines correspond to staircases with a spin jump of $\Delta S=3$ for $\Delta/\delta=0.9$ (top) and $\Delta/\delta=2.3$ (bottom).}\vspace{-0.3cm}
\end{figure}
\section{Conclusion}
We have shown that there exists a small region in the ground-state phase diagram  of a small metallic grain in which pairing correlations and ferromagnetism coexist.
This coexistence regime becomes larger (in the $J_s/\delta$--$\Delta/\delta$ plane) and therefore more accessible to experiments in the presence of a finite Zeeman field.  In particular, we propose that for a given exchange constant (determined by the material used), spin jumps can be observed by tuning a Zeeman field. We have also shown that a quantitative study of the intermediate regime requires the use of the exact solution. Furthermore, the mean-field approximation is qualitatively different in that it does not predict any pairing correlations in the intermediate regime of partial spin polarization. 

In this work, we have ignored mesoscopic fluctuations and focused on a grain with an equidistant single-particle spectrum. It would be interesting to study how mesoscopic fluctuations affect the boundaries of the intermediate phase and the size of spin jumps.

We thank L. Fang, S. Girvin, S. Rombouts, S. Rotter, R. Shankar and A.D. Stone for useful discussions. K. Van Houcke acknowledges financial support of the Fund for Scientific Research - Flanders (Belgium), and the hospitality of the Center for Theoretical Physics at Yale University where this work was completed.  This work was supported in part by the U.S. DOE grant No. DE-FG-0291-ER-40608.

\vfill
\end{document}